\documentclass[twocolumn,seceq]{jpsj2} 
\def\Journal #1,#2,#3,#4#5#6#7{#1 {\bf #2} (#4#5#6#7) #3}
\def\ggs{\buildrel\textstyle > \over {\hbox{\raise0.2ex\hbox{$\sim$}}}}
\def\lls{\buildrel\textstyle < \over {\hbox{\raise0.2ex\hbox{$\sim$}}}}
\def\gsim{\,\lower0.75ex\hbox{$\ggs$}\,}
\def\lsim{\,\lower0.75ex\hbox{$\lls$}\,}
\title
{Thermopower of a Quantum Dot in a Coherent Region}
\author
{\textsc{Takeshi NAKANISHI}$^{1,2}$ and \textsc{Takeo KATO}$^2$}
\inst{${\ }^1$ National Institute of Advanced Industrial Science and Technology\\
 1 Umezono, Tsukuba 305-8568, Japan\\
and CREST, JST\\
 4-1-8 Hon-machi, Kawaguchi, Saitama 332-0012 Japan\\
${\ }^2$Institute for Solid State Physics, The University of Tokyo\\
5-1-5 Kashiwanoha, Kashiwa, Chiba 277-8581, Japan\\
}
\abst{
Thermoelectric power due to coherent electron transmission through a quantum dot
is theoretically studied. 
In addition to the known features related to resonant peaks, we show that a novel significant
structure appears between the peaks. 
This structure arises from the so-called transmission zero, which is characteristic of coherent transmission through several quantum levels.
Because of sensitivity to phase-breaking effect in quantum dots, this novel structure 
indicates degree of coherency in the electron transmission.
}
\kword{thermopower, quantum dot, coherency, Aharonov-Bohm effects, conductance, Landauer formula, Mott formula}
\begin{document}
\maketitle
\section{Introduction}
Understanding of electron transport though quantum dots (QDs) has been of great importance in recent development of quantum-state engineering in semiconductor heterostructures.
Although most experiments have focused on conductance, thermoelectric power (TEP) also gives useful information about the transport processes through QDs.
The sequential tunneling theory predicts a sawtooth-shaped TEP oscillation at high temperatures in a Coulomb blockade regime,~\cite{Beenakker_and_Staring_1992a} while co-tunneling processes are expected to suppress TEP between Coulomb blockade peaks at low temperatures.~\cite{Turek_and_Matveev_2002a}
These predictions have been confirmed in recent experiments of QDs fabricated in two-dimensional electron gases~\cite{van_Houten_et_al_1992a,Dzurak_et_al_1997a,Scheibner_et_al_2006a} and single-wall carbon nanotubes.~\cite{Small_et_al_2003a,Llaguno_et_al_2004a}
In the previous theoretical studies of TEP, only incoherent tunneling processes have been considered.
Therefore, it remains an unsolved problem how coherent electron transmission affects the TEP oscillations.

Coherent electron transport through QDs has first been revealed by conductance measurement of a QD embedded in an Aharonov-Bohm (AB) interferometer.~\cite{Yacoby_et_al_1995a,Yacoby_et_al_1996a,Schuster_et_al_1997a} 
In the experiments, it has been shown that a transmission phase of electrons changes by $\pi$ at each resonant peak in accordance with a Breit-Wigner model.
This indicates that a large amount of electrons retains their coherency during the transmission through QDs. 

Here, let us focus on one important feature in the experiments of the transmission phase.~\cite{Yacoby_et_al_1995a,Yacoby_et_al_1996a,Schuster_et_al_1997a} 
A surprising and unexpected finding in these experiments is that the phase becomes the same between adjacent peaks. 
This indicates that the transmission phase has to change by $\pi$ also at another point between the peaks, 
even though conductance shows no detectable feature there.
In order to explain this intriguing phenomenon called a `phase lapse', a substantial body of theoretical work has been presented.~\cite{Hackenbroich_and_weidenmuller_1997a,Wu_et_al_1998a,Lee_1999a,Taniguchi_and_Buttiker_1999a,Yeyati_and_Buttiker_2000a,Aharony_et_al_2002}
One of the key ideas has been proposed to understand the phase lapse by Lee.~\cite{Lee_1999a}
By general discussion based on the Friedel sum rule, he has shown that identical vanishing of the transmission
amplitude, called a transmission zero, may occur at a specific energy in quasi-1D systems with the time reversal symmetry.
He has claimed that the abrupt jump of the transmission phase 
originates from this transmission zero.
The existence of the transmission zero has been confirmed in simple non-interacting models.~\cite{Taniguchi_and_Buttiker_1999a,Yeyati_and_Buttiker_2000a,Silva_et_al_2002a}
Recently, it has been shown that the transmission zeros survive even in the presence of Coulomb interaction 
within the Hartree approximation.~\cite{Golosov_and_Gefen_2006a} 

In this paper, we study TEP due to coherent electron transmission through a QD by
a non-interacting model. We show that, in addition to the known TEP oscillation,
a novel significant structure appears at the transmission zero, while
no clear feature is observed in conductance there. The condition for appearance of
this structure is discussed in the multi-level QD systems. We also show that this novel structure is 
suppressed by weak phase-breaking of electrons in QDs.
These features provide us useful information about coherency of electrons transferring through the QD.

The outline of this paper is as follows. In \S\ref{sec:formulation}, we formulate 
TEP of a QD based on the Landauer formula. An artificial lead is also introduced to describe 
phase breaking of electrons in the QD. In \S\ref{sec:result}, we calculate TEP as a function of the 
chemical potential, and discuss the characteristic structures near the transmission zeros. 
Finally, the results are summarized in \S\ref{sec:summary}.

\section{Thermoelectric Power of a Quantum Dot}
\label{sec:formulation}
\subsection{Formulation of Thermoelectric Power}
In a coherent regime, conductance and TEP of mesoscopic systems are given by the Landauer formula,~\cite{van_Houten_et_al_1992a,Landauer_1957a,Buttiker_1986a,Buttiker_1988a,Sivan_and_Imry_1986a,Streda_1989a, Butcher_1990a} 
\begin{eqnarray}
G(\mu, T)&\!=\!&{e^2\over\pi\hbar}\int d\varepsilon T(\varepsilon)\left[-{\partial f\over \partial \varepsilon}\right],
\label{Eq: Landauer formula} \\
S(\mu, T)&\!=\!&-{1 \over eT}{
\int d\varepsilon T(\varepsilon)(\varepsilon\!-\!\mu)[-{\partial f/ \partial \varepsilon}]
\over
\int d\varepsilon T(\varepsilon)[-{\partial f/ \partial \varepsilon}]},
\label{Eq: TEP}
\end{eqnarray}
with a transmission probability $T(\varepsilon)$ and a chemical potential $\mu$ in leads.
Here, the derivative of a Fermi distribution function $f$ is given by $-(\partial f/\partial \varepsilon)\!=\!(4k_B T)^{\!-\!1}\cosh^{\!-\!2} ((\varepsilon\!-\!\mu)/2k_B T)$.
TEP is rewritten as $S\!=\!-\langle \xi \rangle/(eT)$, where $\langle \xi \rangle$ is an average of an internal energy $\xi \!=\! \varepsilon \!-\! \mu$. 
Hence, TEP can be interpreted as a measure of an asymmetry in the transmission probability $T(\varepsilon)$ near the Fermi energy in the range of the thermal broadening $k_B T$. 
Here, we should note that the enhancement of TEPs is expected around transmission zeros ($T(\varepsilon)=0$), because the denominator of eq.~(\ref{Eq: TEP}) becomes vanishingly small there at low temperatures.
Throughout this paper, the exact expression (\ref{Eq: TEP}) is used for the calculation of TEPs. 

Here, we comment on the Mott formula.~\cite{Cutler_and_Mott_1969a} 
The Mott formula has been used widely for analysis of the TEP measurements.~\cite{van_Houten_et_al_1992a,Dzurak_et_al_1997a,Scheibner_et_al_2006a,Small_et_al_2003a,Llaguno_et_al_2004a}
It is derived by the Sommerfeld expansion (\ref{Eq: Landauer formula}) and (\ref{Eq: TEP}) with taking up to the first order of $T$ as
\begin{equation}
S_{\rm M}(\mu, T)\!=\!-\frac{\pi^2 k_{\rm B}^2 T}{3e}{1\over G(\mu,T)}{\partial G(\mu, T)\over \partial \mu}.
\label{mott}
\end{equation}
This formula assumes that the TEP is determined only by the asymmetry of conductance at the Fermi energy.
It is a good approximation as far as the product $T(\varepsilon)[-\partial f/\partial \varepsilon]$ is sufficiently large near the Fermi energy. 
The Mott formula is, however, not applicable for the case that the asymmetry of the product far from the Fermi energy is much greater than the contribution near the Fermi energy.
In the present study, the Mott formula gives correct results at moderate or low temperatures, while it shows clear deviation from the exact result~(\ref{Eq: TEP}) at high temperatures. 
In Appendix we will discuss the validity of the Mott formula in detail and give  a new `extended' Mott formula, which always reproduces the correct TEP
of non-interacting systems with arbitrary  transmission probability $T(\varepsilon)$.

\subsection{Model Hamiltonian}

\begin{figure}[tbp]
\begin{center}
\includegraphics[width=60mm]{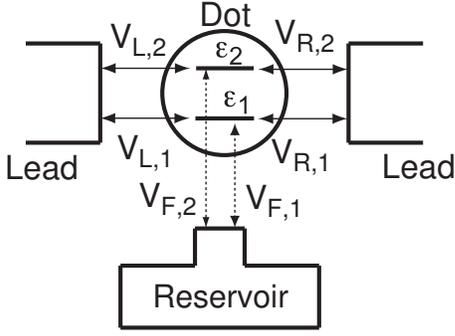}
\caption{
A system composed of two ideal leads and a quantum dot,
and model parameters for the two-level system. Another lead connected to a reservoir
is introduced to describe the phase-breaking effect.
}
\label{Fig: Model}
\end{center}
\end{figure}

We study TEP in a coherent region by the model Hamiltonian
\begin{equation}
H\!=\!\sum \varepsilon_{k,\alpha} C_{k,\alpha}^\dagger C_{k,\alpha}
\!+\!\sum_j \varepsilon_{j} d_{j}^\dagger d_{j}
\!+\!\sum_{k,\alpha,j}[V_{\alpha,j}C^\dagger_{k,\alpha} d_{j}\!+\!H.c.],
\label{Eq:Hamiltonian}
\end{equation}
where operators $c_{k,\alpha}$ refer to electronic states in a
left $(\alpha\!=\!L)$ and right $(\alpha\!=\!R)$ lead, and operators $d_j$ ($j = 1,\cdots,N$) to quantum states in the QD. 
In the presence of the time reversal symmetry, we can take real coupling strengths $\{ V_{\alpha,j} \}$.
The model for $N=2$ is schematically shown in Fig.~\ref{Fig: Model} 
(the role of the reservoir will be explained in the next subsection).
For this non-interacting model, the transmission coefficient can be expressed in terms of the Green's function in a matrix form.~\cite{Silva_et_al_2002a}.
From the transmission coefficient, conductance eq. (\ref{Eq: Landauer formula}) and TEP eq. (\ref{Eq: TEP}) are calculated by numerical integration.

\subsection{Phase-breaking effect}
In reality, there is always some inelastic or phase-breaking scattering.
Partial phase-breaking effect can be introduced by adding one fictitious voltage probe $\alpha\!=\!F$ connected to a reservoir.~\cite{Buttiker_1986a,Buttiker_1988a} 
In the study of the phase-breaking, we only consider the two-level case ($N=2$) with the coupling
$V_{F, 1}\!=\!V'$ and $V_{F, 2}\!=\!V'\exp({\rm i}\theta)$ (see Fig.~\ref{Fig: Model}).
For simplicity, the phase factor $\theta$ is taken as $\pi$.~\cite{footnote1} 
The chemical potential of the reservoir is determined by the condition that 
the current through the fictitious voltage probe vanishes. Then, conductance is given by
\begin{equation}
G\!=\!{e^2\over \pi\hbar} \left[T_{RL}
\!+\!{T_{RF} T_{LF} \over T_{RF}\!+\!T_{LF}}\right],
\end{equation}
with
\begin{equation}
T_{\alpha\alpha'}\!=\!\int d\varepsilon T_{\alpha\alpha'}(\varepsilon)\left[-{\partial f\over \partial \varepsilon}\right],
\end{equation}
where $T_{\alpha\alpha'}(\varepsilon)$ is the transmission 
probability from a lead $\alpha'$ to $\alpha$.
For symmetric case $T_{LF}(\varepsilon)\!=\!T_{RF}(\varepsilon)$,
TEP is easily calculated with the effective transmission probability 
$T_{RL}\!+\!T_{LF}/2$ in eq.\ (\ref{Eq: TEP}).
Although phase breaking is considered by the minimal model, which should be improved for quantitative comparison with
experiments, we can examine the crossover from a fully coherent region to an incoherent region.
\par
\section{Results}
\label{sec:result}
\subsection{Two-level case}
\begin{figure}[tbp]
\begin{center}
\includegraphics[width=75mm]{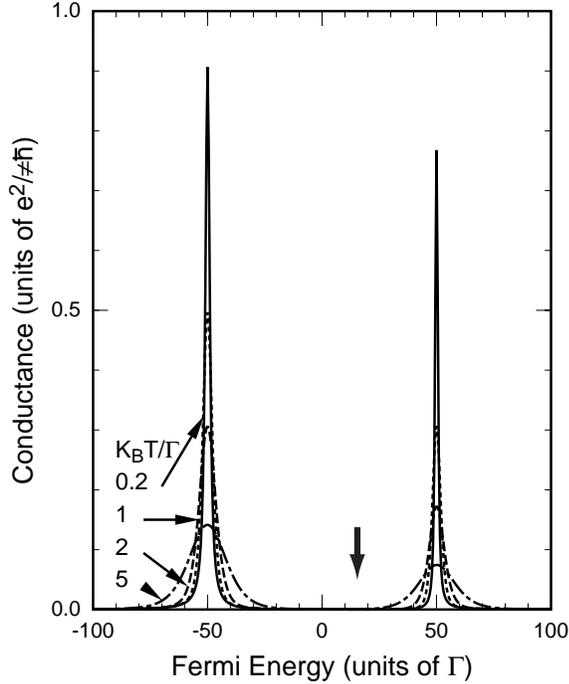}
\caption{Conductance with resonant peaks at $\varepsilon_1\!=\!-50\Gamma$ and $\varepsilon_2\!=\!50\Gamma$ for several temperatures, $k_B T/\Gamma\!=\!0.2$ (solid line), $1$ (dotted line), $2$ (dashed line), and $5$ (dot-dashed line). The vertical arrow indicates the
transmission zero.}
\label{Fig: Fano Conductance}
\end{center}
\end{figure}
\begin{figure}[tbp]
\begin{center}
\includegraphics[width=75mm]{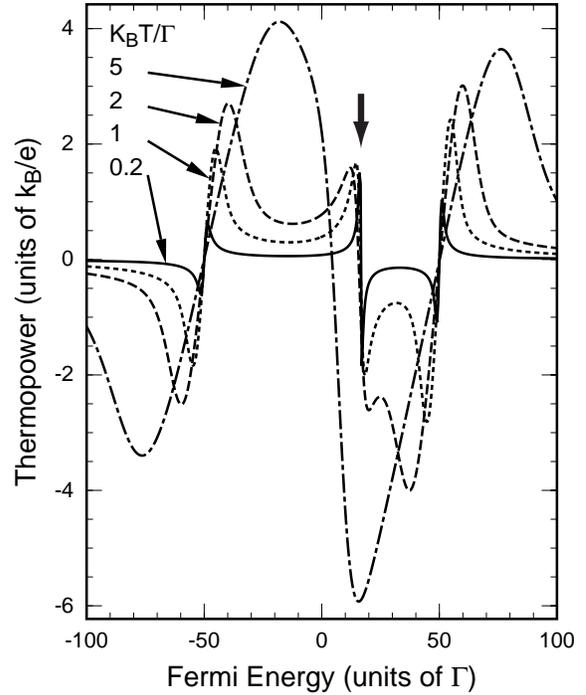}
\caption{Thermoelectric power for the same parameter set as used in the conductance of 
Fig.~\ref{Fig: Fano Conductance}.
The vertical arrow indicates the transmission zero.}
\label{Fig: Fano}
\end{center}
\end{figure}
We start with the two-level case ($N=2$) for the symmetric coupling $V_{\alpha,j} = V_j$ ($j=1, 2$)
in the absence of the phase-breaking effect. The transmission coefficient is calculated as
\begin{equation}
t(\varepsilon)\!=\!{\Gamma_1\!+\!\Gamma_2\over D}(\varepsilon\!-\!\varepsilon_0),
\label{Eq:Transmission_Dot}
\end{equation}
where
\begin{equation}
D \!=\!(\varepsilon\!-\!\varepsilon_1)(\varepsilon\!-\!\varepsilon_2)
\!+\!{\rm i} (\Gamma_1\!+\!\Gamma_2)(\varepsilon\!-\!\varepsilon_0),
\label{Eq:Level_Width_and_Shift}
\end{equation}
with $\Gamma_i \!=\! 2\pi |V_i|^2 \rho$ and the density of states $\rho$ in the leads.
The transmission probability $T(\varepsilon) = |t(\varepsilon)|^2$ vanishes at a specific energy $\varepsilon_0\!=\!(\Gamma_1\varepsilon_2\!+\!\Gamma_2\varepsilon_1) / (\Gamma_1\!+\!\Gamma_2)$ between the resonant peaks at $\varepsilon = \varepsilon_1$ and $\varepsilon_2$. 
This point is called a transmission zero.
\par
%
Figure~\ref{Fig: Fano Conductance} shows conductance calculated for $\Gamma_1\!=\!2\Gamma_2\!=\!\Gamma$ at several temperatures.
Two resonant peaks with the Breit-Wigner line-shape are shown at $\varepsilon_1\!=\!-50\Gamma$ and $\varepsilon_2\!=\!50\Gamma$. 
Although the transmission zero is located at $\mu\!=\!\varepsilon_0\!=\!50\Gamma/3$ (indicated by the arrow in the figure), it is not clearly seen in conductance.
\par
TEP calculated for the same parameter set is shown in Fig.~\ref{Fig: Fano}.
Around the conductance peaks, $\varepsilon = - 50 \Gamma$ and $\varepsilon = 50 \Gamma$,
TEP shows a linear dependence on the Fermi energy with a slope $dS/d\mu=1/(eT)$. 
At moderate or low temperatures, TEP between the conductance peaks
is suppressed from this linear dependence.
These features can be understood by the theory based on sequential-tunneling 
and co-tunneling~\cite{Beenakker_and_Staring_1992a,Turek_and_Matveev_2002a}
as explained in \S\ref{sec:phase_breaking}.
The most characteristic finding is the additional significant structure around the transmission zero 
($\mu\!=\!\varepsilon_0\!=\!50\Gamma/3$, indicated by the arrow in the figure). We can relate this novel structure to the vanishing transmission
amplitude as follows. By expanding the transmission amplitude around the transmission zero by $\varepsilon - \varepsilon_0$, and by
using the Mott formula justified at low temperatures, the TEP is approximately obtained as
\begin{equation}
S_{\rm M}(\mu,T)\!\approx\!-{\pi^2 k_B^2 T\over 3e}{2(\mu\!-\!\varepsilon_0) \over
(\mu\!-\!\varepsilon_0)^2 \!+\!\pi^2 k_B^2 T^2/3}.
\label{Eq:Two level Thermopower}
\end{equation}
This form fits the result shown in Fig.~\ref{Fig: Fano} at low temperatures.
From eq.~(\ref{Eq:Two level Thermopower}), it can be shown that 
TEP takes a maximum and minimum value at $\mu = \varepsilon_0
\mp \pi k_{\rm B} T/\sqrt{3}$ as
\begin{equation}
S_{\rm Max}\!\approx\!\pm \pi k_B/\sqrt{3} e,
\label{Eq: TEP Max}
\end{equation}
in the low-temperature limit. For high temperature $\Gamma\!\ll\!k_{\rm B}T\!\ll\!\Delta\equiv \varepsilon_2\!-\!\varepsilon_1$, one observes a sawtooth-like shape as predict by the sequential tunneling theory.~\cite{Beenakker_and_Staring_1992a}
The interference effects responsible for the transmission zero are smeared out due to the thermal broadening, and
TEP vanishes at the middle point between $\varepsilon_1$ and $\varepsilon_2$ regardless of the zero transmission.
\par
Next, we replace the coupling for the level 2 by an asymmetric one
($V_{\rm R,2} = -V_{\rm L,2}=V_2$) leaving the level 1 symmetric 
($V_{\rm L,1}=V_{\rm R,1}=V_1$). Then, the transmission
coefficient becomes the difference of two Breit-Wigner line-shapes as
\begin{equation}
t(\varepsilon)\!=\! \frac{\Gamma_1}{\varepsilon - \varepsilon_1 + {\rm i}\Gamma_1}-\frac{\Gamma_2}{\varepsilon - \varepsilon_2 + {\rm i}\Gamma_2},
\end{equation}
and therefore the transmission amplitude never vanishes.
Reflecting the disappearance of the transmission zero, the TEP has no significant structure
between the resonant peaks.
Thus, the appearance of the novel structure in TEP between the resonant peaks 
is related to the sign of the couplings.
\par
%
In general, the transmission probability vanishes between $j$-th and $(j\!+\!1)$-th conductance peak for the case that the relative coupling sign, $\sigma_{j}\!\equiv\!\mbox{sign}(V_{L,j}V_{R,j}V_{L,j+1}V_{R,j+1})$, equals $\!+\!1$, while no transmission zero appears for $\sigma_{j}\!=\!-1$.~\cite{Silva_et_al_2002a,Taniguchi_and_Buttiker_1999a,Yeyati_and_Buttiker_2000a} 
Hence, the appearance of the novel structure between the $j$-th and $(j+1)$-th conductance peak depends only on the relative coupling sign $\sigma_{j}$. 
We demonstrate this by studying the multi-level case in the next subsection.
\par
%
\subsection{Multi-level case}
\label{sec: multi}
%
\begin{figure}[tbp]
\begin{center}
\includegraphics[width=75mm]{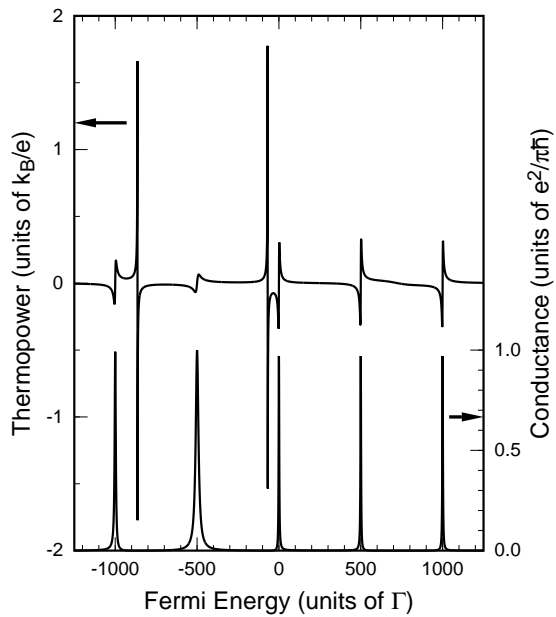}
\caption{
Conductance and thermoelectric power of a dot with five quantum levels coupled
to leads with individual coupling strengths for $k_B T=0.2\Gamma$.}
\label{Fig: Consecutive Peaks}
\end{center}
\end{figure}
Figure~\ref{Fig: Consecutive Peaks} shows an example of the multi-level case at low temperature $k_B T=0.2\Gamma$.
The conductance peaks at $\varepsilon_j/\Gamma=500(j-3)$ for $j=1,\dots,5$ are corresponding to the levels in the QD with individual couplings $V_{\alpha,1}=\sqrt{2}V$, $V_{\alpha,2}= \sqrt{5}V$, and $V_{\alpha,3}= V_{\alpha,5}= V$, while $V_{L,4}=V$ and $V_{R,4}= -V$, 
where $\Gamma \!=\! 2\pi |V|^2 \rho$. 
TEP for the same parameter set shows, in addition to small spikes corresponding to the conductance peaks, the significant structures with an amplitude $\sim (\pi/\sqrt{3})(k_B/e)$ at the transmission zeros, $\varepsilon \sim -860\Gamma$ and $-80\Gamma$. 
These structures are observed only among the conductance peaks $j = 1, 2$ and $3$ because the relative coupling signs take $\sigma_{j}\!=\!1$ for $j=1,2$. 
On the other hand, no significant structure is found among conductance peaks $j=3,4$ and $5$, because $\sigma_{j}\!=\!-1$ for $j=3, 4$.
Thus, the appearance of the significant structures between resonant peaks can be 
related to the relative coupling signs, which gives the phase information of wavefunction in the QD. 

In general, observation of the transmission zeros in conductance measurement is rather difficult. 
For example in the ordinal lead-dot-lead configuration, it is difficult to identify the transmission zeros in the region where the conductance is exponentially suppressed far from the conductance peaks. 
In principle, in a hybrid structure with an AB ring, the zero transmission can be detected by the abrupt jump of the transmission phase.~\cite{Lee_1999a} 
The actual analysis for this type of hybrid systems, however, is highly complicated, since the whole system consisting of a reference arm and quantum dot should be considered as one resonator;
For example, interference effect within the AB ring significantly affects conductance.~\cite{Wu_et_al_1998a,Aharony_et_al_2002,Nakanishi_et_al_2004a} 
Compared with conductance measurement, observation of the transmission zeros by 
TEP may have an advantage because both measurement and analysis are simple.
\par
\subsection{Phase-breaking effect}
\label{sec:phase_breaking}
\begin{figure}[tbp]
\begin{center}
\includegraphics[width=75mm]{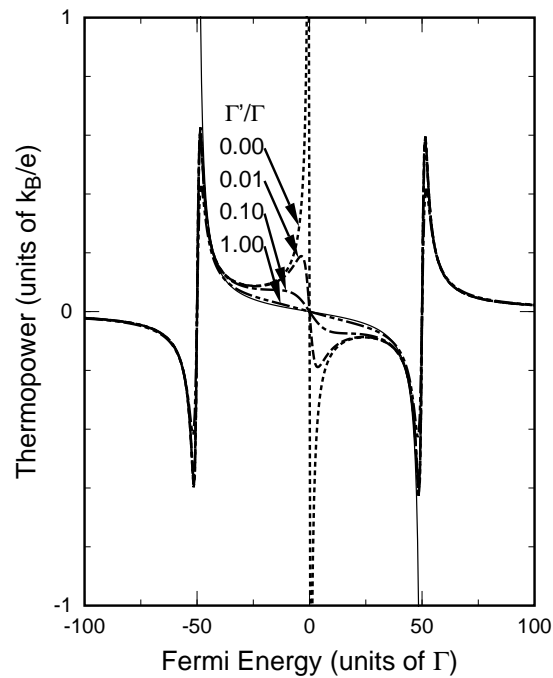}
\caption{The phase-breaking effect on thermoelectric power at low temperature $k_B T\!=\!0.2\Gamma$ for the several
phase-breaking strengths $\Gamma'/\Gamma\!=\!0$ (dotted line), $0.01$ (dashed line), $0.1$ (dot-dashed line) and $1$ (dot-dot-dashed line).
Solid thin line shows the thermoelectric power predicted by the co-tunneling theory.}
\label{Fig: dephasing}
\end{center}
\end{figure}
Let us move on the phase-breaking effect on TEP. 
The strength of phase breaking can be controlled by the coupling $\Gamma'\!=\!2\pi|V'|^2\rho'$ with the density of states $\rho'$ in the fictitious probe.
Figure \ref{Fig: dephasing} shows the phase-breaking effect on TEP of a QD with two quantum levels ($N=2$) for several values of $\Gamma'$, where we chose $\Gamma_1\!=\!\Gamma_2\!=\!\Gamma$, $\varepsilon_1\!=\!-50\Gamma$ and $\varepsilon_2\!=\!50\Gamma$ as a typical example.
With the increase of phase breaking, the structure at the transmission zero ($\mu\!=\varepsilon_0\!=\!0$) is suppressed, because the destructive interference between two possible paths through the two levels in the QD, which is responsible for the transmission zero, is sensitively diminished by the loss of coherency. 
On the other hand, the small spikes corresponding to the conductance peaks is slightly affected by the phase breaking, because the conductance peak, which is determined by one dominant path through one quantum level, is insensitive to the loss of coherency as far as $\Gamma'\ll \Gamma$.
\par
For large phase-breaking, it is expected that the higher-order processes with respect to
the coupling $V_{\alpha, j}$ are terminated, and that only the lowest-order contribution 
representing the sequential-tunneling and/or co-tunneling process survives. 
Then, the calculation in a coherent region can be related to the known theory based on these two tunneling processes as follows.~\cite{Beenakker_and_Staring_1992a,Turek_and_Matveev_2002a} 
Far from the resonant peaks, TEP is determined dominantly by the inelastic co-tunneling process, which predicts the energy-dependence of TEP as~\cite{Turek_and_Matveev_2002a}
\begin{equation}
S_{\rm co}\!=\!{k_B^2 T \over e}{4\pi^2\over 5} \left({1\over \mu\!-\!\varepsilon_1}\!+\!{1\over \mu\!-\!\varepsilon_2}\right).
\label{Eq: cotunneling}
\end{equation}
This expression, which is drawn by the solid thin line in Fig.~\ref{Fig: dephasing}, well explains the behavior of TEP far from the conductance peaks for large phase breaking. 
The co-tunneling theory breaks down near the conductance peaks, where the sequential tunneling process becomes dominant.~\cite{Turek_and_Matveev_2002a}
Around the conductance peaks, the slope ${\rm d}S/{\rm d}\mu = 1/(eT)$ is predicted from the sequential tunneling theory~\cite{Beenakker_and_Staring_1992a} in the quantum limit $U \ll \Delta$, where $U$ and $\Delta$ are a Coulomb interaction in the QD and a level spacing, respectively.~\cite{footnote2} 
This predicted slope near the resonant peaks is consistent with all the results of TEP
for weak or moderate phase-breaking. For large phase-breaking ($\Gamma' \gsim \Gamma$), 
the slope becomes smaller than $1/(eT)$ due to broadening of the conductance peaks.
\par
\begin{figure}[tbp]
\begin{center}
\includegraphics[width=75mm]{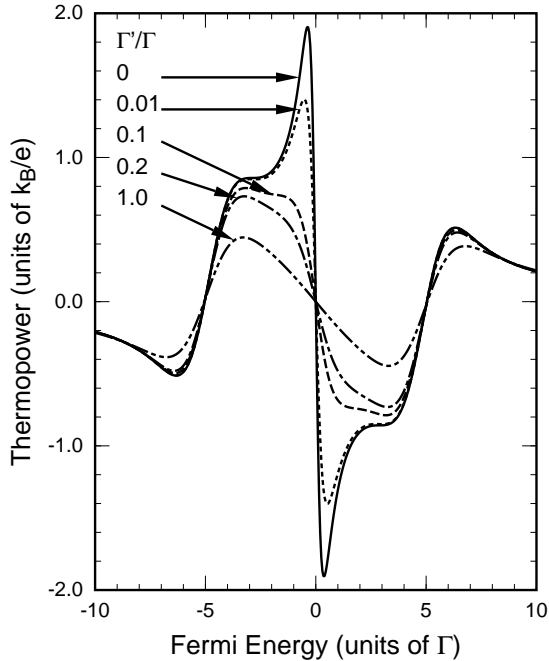}
\caption{The phase-breaking effect on thermoelectric power for a small level spacing $\Delta\!=\!10\Gamma$ ($\varepsilon_1\!=\!-5\Gamma$ and $\varepsilon_2\!=\!5\Gamma$)
at low temperature $k_B T\!=\!0.2\Gamma$.
}
\label{Fig: Delta 10}
\end{center}
\end{figure}
The correspondence with the co-tunneling theory discussed above becomes unclear for a small level spacing $\Delta$, where the higher-order processes are more important.
We show another example of the TEP for a small level spacing $\Delta\!=\!10\Gamma$  ($\varepsilon_1\!=\!-5\Gamma$ and $\varepsilon_2\!=\!5\Gamma$) in Fig.~\ref{Fig: Delta 10}.
The feature of co-tunneling, which can be seen in strong suppression of TEP 
expressed by (\ref{Eq: cotunneling}), is not found 
even for large phase breaking, and more complicated behavior is observed.
The phase breaking decreases the TEP in a wide range of the Fermi energy
due to suppression of the higher-order tunneling processes.
The structure at the transmission zero is suppressed by phase breaking
more gradually than that for $\Delta\!=\!100\Gamma$ in Fig.~\ref{Fig: dephasing};
Its remnant can be observed for moderate phase-breaking $\Gamma' = 0.2 \Gamma$,
so that the appearance of the transmission zero can be indicated more clearly.
\par
In order to clarify the behaviors of TEP around the transmission zero, we consider the 
perturbation theory
with respect to the coupling $V_{\alpha,j}$. The whole transmission probability including the phase-breaking effect is evaluated
around the transmission zero ($\varepsilon = 0$) up to the forth-order with respect to $V_{\alpha, j}$ as
\begin{equation}
T(\varepsilon) \approx \frac{8 \Gamma \Gamma'}{\Delta^2} + 
\frac{64 \Gamma(\Gamma + \Gamma')}{\Delta^4} \varepsilon^2,
\label{Eq: expansion}
\end{equation}
for $\Gamma, \Gamma' \ll \Delta$. The first term proportional to $\Delta^{-2}$ describes the incoherent 
co-tunneling process, by which the transmission amplitude have a finite value at $\varepsilon = 0$.
The transmission zero in coherent transport appears as a result of the destructive interference between 
two paths, each of which corresponds to the coherent transmission through each level in the QD. 
This interference process start with the forth-order perturbation with respect to $V_{\alpha, j}$.
As a result, the coherent part responsible to the transmission zero appears
in the second term in (\ref{Eq: expansion}), and is proportional to $\Delta^{-4}$. 
From the expansion (\ref{Eq: expansion}), the slope of 
TEP at the transmission zero can be calculated at low temperatures as
%
\begin{equation}
\left. \frac{{\rm d}S_M}{{\rm d}\mu} \right|_{\mu = \varepsilon_0} \approx
- {k_B\over e}\left({16\pi^2 k_B T\over 3\Delta^2}
\left({\Gamma\over\Gamma'}\!+\!1\right)\right),
\label{Eq: slope}
\end{equation}
%
which well approximates the slopes at $\mu=0$ in Figs.~\ref{Fig: dephasing} and \ref{Fig: Delta 10}. 
In the limit $\Gamma' \rightarrow \infty$, the slope agrees with the co-tunneling theory.~\cite{Turek_and_Matveev_2002a}
Within this approximation, the peak height is proportional to 
$(k_{\rm B}/e) (k_{\rm B}T/\Delta) (1+\Gamma/\Gamma')^{1/2}$ at low temperatures.
Hence, a small value of $\Delta$ is suitable to observe the significant structure at the transmission zero.
\par
Thus, the structure of TEP around the transmission zero is sensitive to the phase-breaking effect. 
Therefore, 
TEP may provide a useful tool to measure coherency of electrons transferring through QDs. 
So far, coherency of transmission electrons has been measured by conductance of a QD embedded in an AB ring with a magnetic field. 
The analysis of this system is, however, complicated as discussed in \S\ref{sec: multi}. 
The TEP measurement, which do not need either a hybrid structure like an AB ring or an external magnetic field, may give an alternative
simple method for study of coherency in electron transport, in particular, for the off-resonant region far from the conductance peaks.
\par
\section{Summary}
\label{sec:summary}
Thermoelectric power in a fully coherent region has been studied theoretically. 
It has been shown that a significant structure appears
at the so-called transmission zero, at which a transmission amplitude vanishes.
The appearance of these structures is directly related to the sign of the coupling with leads,
reflecting phase information of wavefunctions in quantum dots. It has also been shown that
these structures are sensitively suppressed by weak phase-breaking, and that the calculated thermoelectric power coincides with the co-tunneling theory for sufficiently large phase-breaking.
It has been proposed that, due to sensitivity to phase breaking, thermoelectric
power can be used to measure electron coherency in a quantum dot, even in case
the Aharonov-Bohm oscillation cannot be used by a fairly small amplitude.
The effects of Coulomb interaction on the thermoelectric power remains an important problem for future study.

\acknowledgement

We acknowledge to Y. Hamamoto for useful discussions about the derivation of the extended Mott formula.

\appendix
\section{Extended Mott formula}
\label{Sec: Extended Mott formula}
The Mott formula is not applicable for the case that the asymmetry far from the Fermi energy 
is much greater than the contribution near the Fermi energy.
Actually, TEP calculated from the Mott formula (\ref{mott}) clearly deviates from the exact formula
(\ref{Eq: TEP}) at high temperature $\Gamma\!\ll\!k_{\rm B}T\!\ll\!\Delta$ in the present study.
In order to understand the deviation, we replace the transmission amplitude by a simple form $T(\varepsilon) = \sum_n \delta(\varepsilon-\varepsilon_n)$, where $\varepsilon_n$ is an energy
level in the QD.
TEP is calculated from (\ref{Eq: TEP}) as $S(\mu,T) = k_{\rm B}\Delta_{\rm min}/eT$, 
where $\Delta_{\rm min} = \varepsilon - \varepsilon_{n_{\rm min}}$ and $n_{\rm min}$ 
is the level index minimizing $|\varepsilon - \varepsilon_n|$.
On the other hand, the Mott formula gives an incorrect result $(\pi^2/3) \tanh(\Delta_{\rm min}/2T)$.
Another example is a point contact almost pinched off.~\cite{Lunde_and_Flensberg_2005a}
For this case, the transmission probability is given by the step function as $T(\varepsilon)\!=\! \Theta(\varepsilon)$. 
TEP is calculated from (\ref{Eq: TEP}) as $S\!=\!(k_{\rm B}/e) \left[-\beta \mu\!+\! \ln(1\!-\!f(0))/f(0) \right]$, 
while the Mott formula (\ref{mott}) gives a different result $S_{\rm M} \!=\! (\pi^2/3) (k_{\rm B}/e) (1-f(0))$,
where $f(0) \!=\! 1/(e^{\!-\! \beta \mu}\!+\!1)$. Thus, we should use the Mott formula carefully by noting its limitation.

For general non-interacting models, we can derive an exact formula
from (\ref{Eq: Landauer formula}) and (\ref{Eq: TEP}) as
\begin{equation}
S(\mu, T)\!=\!-{1\over e}{1\over G(\mu, T)}\int^{\mu} d\mu' {
\partial G(\mu', T)
\over
\partial T}.
\label{Eq: extended Mott}
\end{equation}
This `extended Mott formula' relates TEP to the derivative of conductance 
with respect to the temperature $T$ instead of the chemical potential $\mu$. 
The same relation can be rewritten in a derivative expression as
%
\begin{equation}
{\partial \over \partial \mu}(S(\mu, T)G(\mu, T))=- {1\over e}{\partial G(\mu, T)
\over \partial T}.
\label{Eq: extended Mott}
\end{equation}
%
Since this formula is always correct for non-interacting systems, it will be useful for analysis of experimental results.
On the other hand, it may become invalid in interacting electron systems. 
For example, the transmission probability of a carbon nanotube may depend also on the chemical potential $\mu$ when the Schottky barrier is formed at the interface between the carbon nanotube and leads. Then, a deviation from the formula (\ref{Eq: extended Mott}) will be observed.

%

\begin{thebibliography}{l}
\bibitem{footnote1}The phase factor $\theta$ only changes the effective phase-breaking strength.
However, we have to set it a nonzero value because
the phase breaking effect disappears for $\theta = 0$ in this model.
\bibitem{footnote2}We should note that, in the experiments, the opposite limit ($U \gg \Delta$) is usually 
realized. For this limit, we obtain the slope ${\rm d}S/{\rm d}\mu = 1/(2eT)$, which differs by 
the factor $2$ from the quantum limit $U \ll \Delta$.
\end{thebibliography}
\end{document}